# Coherence effects in LIPSS formation on silicon wafers upon picosecond laser pulse irradiations

Inam Mirza[1], Juraj Sládek[1], Yoann Levy[1,*], Alexander V. Bulgakov[1], Vasilis Dimitriou[2,3], Helen Papadaki[2,3], Evaggelos Kaselouris[2,3], Paulius Gecys[4], Gediminas Račiukaitis[4], Nadezhda M. Bulgakova[1,*]

[1] HiLASE Centre, FZU - Institute of Physics of the Czech Academy of Sciences, Dolní Břežany, Czech Republic
[2] Institute of Plasma Physics and Lasers-IPPL, University Research and Innovation Centre, Hellenic Mediterranean University, 74150, Rethymno, Greece
[3] Physical Acoustics and Optoacoustics Laboratory, Department of Music Technology and Acoustics, Hellenic Mediterranean University, 74133, Rethymnon, Greece
[4] FTMC - Center for Physical Sciences and Technology, Vilnius, Lithuania
*Corresponding authors
E-mail: levy@fzu.cz, bulgakova@fzu.cz

**Abstract.** Using different laser irradiation patterns to modify of silicon surface, it has been demonstrated that, at rather small overlapping between irradiation spots, highly regular laser-induced periodic surface structures (LIPSS) can be produced already starting from the second laser pulse, provided that polarization direction coincides with the scanning direction. If the laser irradiation spot is shifted from the previous one perpendicular to light polarization, LIPSS are not formed even after many pulses. This coherence effect is explained by a three-wave interference, - surface electromagnetic waves (SEWs) generated within the irradiated spot, SEWs scattered from the crater edge formed by the previous laser pulse, and the incoming laser pulse, - providing conditions for amplification of the periodic light-absorption pattern. To study possible consequences of SEW scattering from the laser-modified regions, where the refractive index can change due to material melting, amorphization, and the residual stress formed by previous laser pulses, hydrodynamic modelling and simulations have been performed within the melting regime. The simulations show that stress and vertical displacement could be amplified upon laser scanning. Both mechanisms, three-wave interference and stress accumulation, could enable an additional degree of controlling surface structuring.

## 1. Introduction

Large-area texturing of material surfaces by laser-induced periodic surface structures (LIPSS) has been investigated for decades on various types of substances [1-6]. Fabricating LIPSS with ultrashort laser pulses could represent a cheap single-step way to tune properties of surfaces of any material, such as wettability [7,8], colorizing [9,10], or tribology [11,12] which still makes the study of their formation and self-replication of great importance and interest [13-16]. In the case of low spatial frequency LIPSS (LSFL), the self-replication of structures by laser beam scanning over the surface [4,17,18] makes it possible to cover large areas with diverse degrees of homogeneity depending on laser scanning parameters [19-21]. Besides homogeneity, achieving the LIPSS with high regularity is an essential topic in the investigations related to large-area texturing of surfaces [22-26].

Several conditions have been identified to play a significant role in the regularity of LSFL. Light polarization, if oriented parallel to the scanning direction upon laser processing of the Cr surface, was shown to facilitate a better regularity as compared to perpendicularly polarized light [14]. Irradiation spot size about or smaller than the surface plasmon polariton (SPP) decay length, $L_{SPP}$, was suggested to have a beneficial influence on the LSFL regularity for metals [23,25]. The wavefront curvature [27] was demonstrated to deteriorate the quality of the structures on silicon. Scanning parameters such as pulse overlap, interline overlap, and precise fluence have been pointed out to be of significant influence for allowing regular LSFL formation on silicon surfaces [20,28].

Alongside the mentioned factors, it is important to consider the imprinting mechanisms of the LIPSS for covering large surface areas. During scanning, each laser pulse interacts with a virgin region and an area modified by previous pulses. How the LIPSS replicate themselves to new irradiation areas is important for efficient and good quality nanostructuring of large surface areas at high throughputs. Recently we demonstrated that the LIPSS reproduce themselves in new areas even at slight overlap between irradiation spots [29]. This paper presents a detailed study of the possible mechanisms and processes responsible for LIPSS replication. Different arrangements of the neighbouring irradiation spots at small overlaps between them were considered. This enabled us to make a conclusion that the whole modified region provides a diffraction effect for both initiating and reproducing the LIPSS in new irradiation areas. It has been shown that the light polarization direction relative to the scanning direction plays a key role in LIPSS inscription which is especially demonstrative at small overlaps between pulses. Additionally, the effects of laser-induced stress-wave deformations on the LIPSS generation upon laser scanning over a silicon surface are computationally investigated by numerical simulations using finite element method (FEM) models within the melting regime.

## 2. Experimental

The experiments were carried out using a Yb:KGW pulsed laser (PHAROS, Light Conversion) with a wavelength of 1030 nm and 6 ps pulse duration (FWHM). The beam with the Gaussian spatial and temporal profiles was focused normal to a silicon surface using an f-theta lens (163 mm focal length) from a galvo scanner (SCANLAB SCANcube III 14). The galvo scanner was used to arrange the different patterns of the irradiation spots, as shown in figure 1. It should be noted that the precision of the galvo scanner (about ±2.5 µm on the sample surface) did not allow for a perfect spacing between the spots, and fluctuations in inter-spot distances can be seen in some of the images shown below.

The linearly polarized laser light was attenuated by the combination of a half-waveplate and a polarizing beam splitter. The energy per pulse, $E_p$, was measured by a thermopile-based sensor 12A-P (Ophir Optronics). The pulse repetition rate was 25 and 50 kHz in scanning experiments, which, however, did not influence the final structuring results. The spot diameter at $1/e^2$ of the intensity measured as described in [30] was $2w_0 = 27$ µm, where $w_0$ is the beam waist. The polarization direction on the sample was adjusted via the rotation of a second half-waveplate positioned after the f-theta lens. It was verified by measuring the transmitted power when placing an extra Glan-Taylor polarizer, which was temporarily installed above the processing area and appropriately oriented.

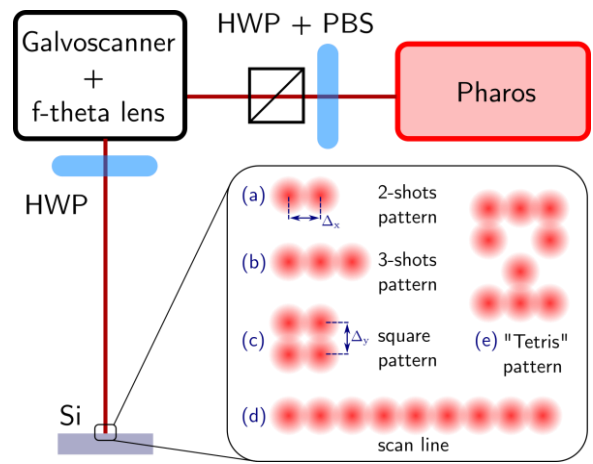

**Figure 1.** Schematics of the experimental setup and some examples of the irradiation patterns (a)–(e). A half-waveplate (HWP) and polarizing beam splitter (PBS) were employed for the laser beam attenuation.

The incidence of the laser beam was normal on the (100) surface plane of a silicon single crystal ~500 µm thick (single-side polished Si wafers from PI-KEM, p-doped with boron at $10^{-10}$ Ω·cm). The laser fluences indicated below are the peak ones calculated as $F_0 = 2E_p/(\pi w_0^2)$. The damage threshold fluence $F_{th}$ was determined to be ~0.65 J/cm². The images presented below were obtained by a differential interference contrast (DIC) optical microscope Olympus BX43, while the crater profiles were analysed with a confocal laser scanning microscope Olympus OLS5000.

## 3. Results and discussion

Figure 2(a) shows a typical modification spot produced by a single laser pulse on the sample surface for a peak fluence of $F_0 = 1.64$ J/cm$^2$. Around the pronounced circular ablation crater with a diameter of ~18 μm and a maximum depth at the spot center of ~0.6 μm (figure 2(b)), an amorphization ring (of slightly lighter shade than the pristine area on the image) can be recognized. Its external boundary is outlined by the blue line, and the boundary of the 1/e$^2$ irradiation spot size is marked by the red line. The amorphization is a usual effect observed at ultrashort laser pulse irradiation of crystalline Si surfaces [31-34]. At this particular peak fluence, the amorphization ring has an external diameter $d(F_0) = (20.9 \pm 0.6)$ μm. Based on this typical laser-modification size, we have investigated the initiation the LIPSS formation upon scanning, involving so-called non-local feedback [18].

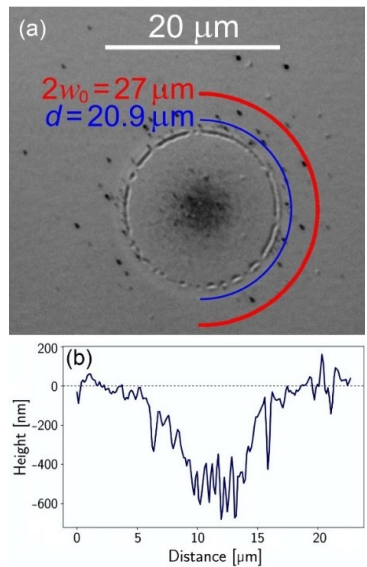

**Figure 2.** (a) DIC microscope image of the spot structure formed by a single laser pulse with 6 ps duration at 1030 nm wavelength. $F_0 \approx$ 1.64 J/cm$^2$. The border of the ablated area is clearly seen as a dark outline. The blue line corresponds to the external boundary of laser-induced amorphization, and the red line marks the irradiation spot size (1/e$^2$). (b) A typical crater profile formed at a single shot regime.

### *3.1 Coherence and polarization effects*

The question arises as to how a laser-modified area can affect the spot created by subsequent laser pulses when the spatial shift between the pulses is of the order of the external diameter of the laser-modified zone $d(F_0)$. To investigate this, the samples were irradiated by several laser pulses while keeping the distances $\Delta x$ and/or $\Delta y$ between them close to $d(F_0)$. With such a distance between the irradiation spot centers, the modification areas barely intersect while the next irradiation spot (1/e$^2$) partially overlaps with the modification produced by the previous pulse. Different patterns of irradiation were applied as outlined in figure 1(a) to (d). The orientation of the laser polarization was varied with respect to the pulse-to-pulse shift in the *x* and/or *y* directions, respectively the horizontal and vertical directions in the images.

In all patterns except (d) in figure 1, created by continuous scanning, each shot was fired one by one by opening the laser pulse picker on demand and with a time delay of about one second or more between subsequent shots. The use of the galvo scanner explains a misalignment of up to 5 μm between some laser-modified spots visible in the images presented below.

Figure 3 shows modifications produced by two pulses (irradiation pattern (a) in figure 1) with different polarization directions relative to the pulse-to-pulse shift. A region between the laser-modified spots brighter than the surrounding area is visible where the amorphization rings of individual pulses overlap, as shown in several studies [28,35]. Well-pronounced LIPSS are formed in the second crater only when the polarization of the second pulse is parallel to the beam displacement direction, irrespectively to the polarization in the first pulse. The formed structures are the LSFL of type I with the orientation perpendicular to the polarization [36]. Their spatial period of $\Lambda = (1033\pm68)$ nm is in accordance with expectations ($\Lambda \sim \lambda$ at normal incidence and for an effective number of pulses per point, $N_{eff}$, close to 1) and with values reported for Si irradiated with similar laser parameters [13,21]. The LSFL are not produced when the polarization of the second shot is perpendicular to the laser beam displacement. Thus, a conclusion could be drawn that the surface scattered electromagnetic waves (SEWs) generated by the second pulse action and propagating preferentially along the polarization direction are scattered at a step-like boundary of the first ablation spot [37,38]. As a result, the SEWs generated on the irradiation spot area interfere with those scattered by the boundary of the previous crater. In turn, this leads to an amplification of the classical mechanism of LIPSS formation, based on which the periodic absorption pattern is created by the interference of the SEWs and the incident laser wave as described in the seminal work by Sipe et al. [39]. Indeed, the curvature of the LSFL in the second spot replicates the curvature of the ablation crater of the first spot (figure 3). Note that, in the case of the absence of a scattering obstacle (previously formed crater) on the way of SEWs, the Sipe mechanism looks to be insufficient for creating periodic structures in the newly formed crater (images on the left in figure 3) and only SEWs scattered from the crater borders produce a strong amplification of this mechanism (images on the right in figure 3).

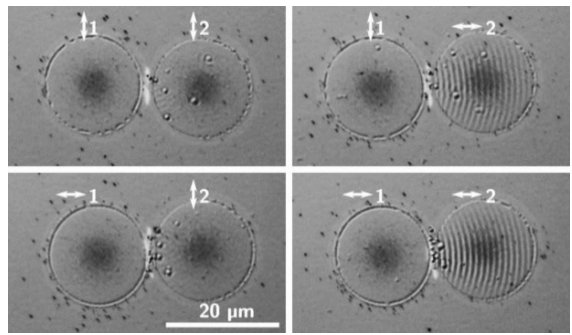

**Figure 3**. DIC microscope images of two-shot irradiation patterns on Si (100) surface in the geometry (a) shown in figure 1. The spot number corresponds to the order of irradiation. Each image was obtained with a different combination of polarization orientations (marked by the two-sided arrows). The peak fluence is 1.64 J/cm$^2$, the pulse duration is 6 ps, and $2w_0 \sim 27$ μm.

To further investigate the distance influence of the borders of the previously produced craters on the formation of LSFL, additional irradiation patterns have been studied. In figure 4, two irradiation geometries are presented: three laser shots in line (figures 4(a)-(d)) and square patterns with two different orientations of polarization, parallel (e) and perpendicular (f) to the pulse-to-pulse shift. It is again evident that the generation of LIPSS by the subsequent laser pulses is not influenced by the polarization of the pulse producing the first modification area. This supports the interpretation that the reason for the LIPSS formation in the subsequent relatively distant spots is the presence of an extended surface defect in the form of the ablation crater boundary surrounded by an amorphous Si ring. If the polarization of the subsequent pulses is parallel to the beam displacement direction, the SEWs are scattered from the boundary of a previously produced crater, resulting in the generation of regular LSFL produced over the next crater area with amplitudes, which look to be amplified from pulse to pulse (figures 4(a) and 4(c)). In the case of laser polarization perpendicular to the beam shift direction, the LIPSS do not replicate the previous crater's curvature over the bottoms of the following craters (figures 4(b) and 4(d)). One can only notice some localized periodic structures that are produced due to the "antenna-like" effect [29] conditioned by the presence of local micro/nano defects (hillocks, holes, deposited particulates) that are seen in figure 4(d).

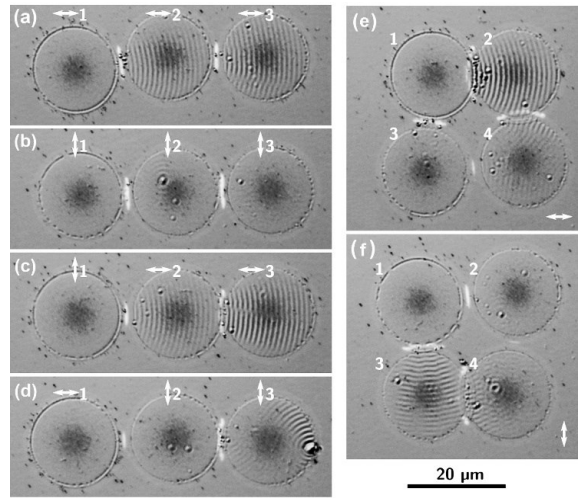

**Figure 4**. DIC microscope images of the laser-produced spots produced in three-shot (a)-(d) and square-pattern irradiation geometries (e) and (f) on Si (100) surface. The laser irradiation conditions are the same as in figure 3. The spot numbering corresponds to the order of irradiation shots. The double-sided arrows indicate polarization orientations. In the square patterns, all pulses had the same polarization, either horizontal (e) or vertical (f).

The examples given in figure 4 demonstrate that the edge of a crater is a determining factor in the formation of LIPSS by the subsequent laser shots in the regime when the craters are not overlapping. It is also seen that the separation between the crater edges should be relatively small, of a few micrometres, or they should touch as in figure 4(e), top for a robust replication of LIPSS. When the edges of the subsequent craters are separated by ca. 3–4 μm and more, the LIPSS replication strongly weakens (see 4$^{th}$ spots in images 4(e) and 4(f)).

Thus, for the LIPSS to form in a crater under the studied configurations, at least two conditions should be fulfilled. First, there should be another crater located sufficiently close (within a distance of a few wavelengths) to induce substantial scattering of the SEWs from its edge. Second, the orientation of the polarization should be appropriately aligned with respect to the edge of the crater produced by the previous pulse. Similar conditions have already been considered in several works. Indeed, significant topography variations result in the formation of LSFL when the polarization vector is perpendicular to a step in the surface relief [37]. Moreover, the distance between the two subsequent pulses plays a key role in the non-local feedback mechanism for LSFL replication [18]. Figure 5, where the "Tetris" intercrater geometries are presented, further confirms these conclusions.

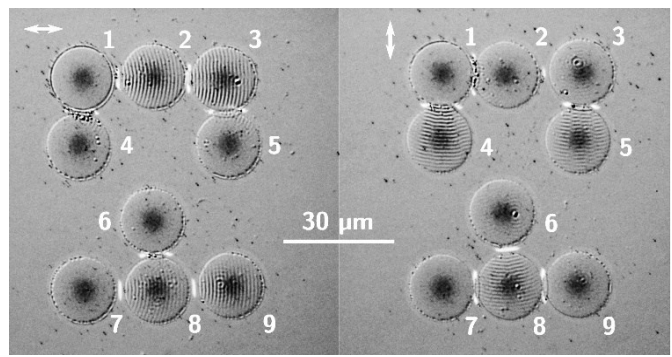

**Figure 5**. DIC microscope images of the laser-produced craters obtained with the same "Tetris" geometry but different polarization directions (left and right panels) on the Si (100) surface. The laser irradiation conditions are the same as in figures 3-4. The spot number corresponds to the order of irradiation shots.

Applying the line-scanning pattern ((d) in figure 1), the formation of highly regular LSFL can be achieved if the laser polarization is parallel to the scanning direction (figure 6, top). The first crater (marked by 1) is free from periodic structures, but they appear already within the second crater. Furthermore, from pulse to pulse, the amplitude of LIPSS is increasing and they lose the initial curvature attributed here to the crater edge. This can be explained by the SEWs scattering not only from the edge of the previous crater but also from the ripples inside it that provide positive pulse-by-pulse feedback, typical for scanning [1,18,36]. If the polarization direction is perpendicular to the scanning direction, no LIPSS are observed except for small diffraction patterns formed on the defects, as seen in figure 6, bottom.

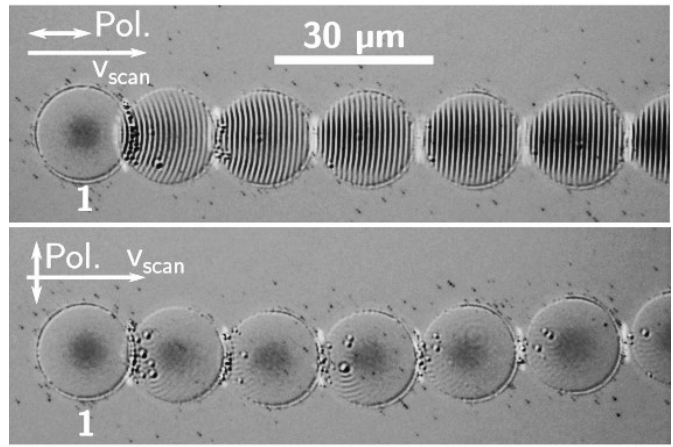

**Figure 6**. DIC microscope images of the laser-produced structures in the case of line-scanning on Si (100) surface with two different polarization orientations ("Pol." with respect to the scanning direction "$v_{scan}$"). The other irradiation parameters are the same as in figures 3–5. The repetition rate was 50 kHz, and the scanning speed was adjusted to get about 20 μm distance between laser spot centers. The first pulse in the line is marked by 1.

The LSFL studied here relate to so-called ablative LIPSS, whose formation can be strongly influenced by post-irradiation thermodynamic processes, material relocation, and hydrodynamic instabilities [40-45]. In the next Section, we discuss the possible effects on LIPSS formation, which can be caused by the interaction of the laser-induced stress waves with the laser modification produced by the previous laser pulse.

*3.2 Exploring the role of hydrodynamics*

The LIPSS formation is a complicated phenomenon that does not end with the periodic light absorption by the surface due to interference of the laser and surface electromagnetic waves, providing higher material vaporization in the hotter regions. It can also involve hydrodynamic material relocations over the surface with a possibility of instability development within the laser-molten pool [42-44], Coulomb explosion [46], and the self-organization mechanism involving non-linear dynamical relaxation of polarization-induced asymmetry of the initial electron kinetic energy distribution [47]. Here, we consider additional factors that may have an impact on the LIPSS formation, namely laser-induced stresses and stress waves propagating over the irradiated target and interacting with surface peculiarities, including those generated by previous laser pulses.

Upon pulsed laser action, strong surface acoustic waves are generated, which can induce nonthermal desorption of atoms, molecules, or atomic clusters from laser-irradiated surfaces and cause surface modification [48,49]. Here, we raise the question of whether the surface stress waves can influence the periodic pattern formation discussed in the previous sections. For this sake, the numerical model developed for nanosecond pulse laser irradiation in the regimes from elastic to melting ones [50-52] has been applied for ultrashort (10 picosecond) laser pulses. Considering that the characteristic electron-lattice thermalization time for silicon is relatively short (~7-8 ps [53,54]), a single-temperature model has been used.

The FEM model is governed by the heat conduction partial differential equation (PDE) coupled to the structural PDE and simulates the stress wave propagation considering the hydrodynamic effects by a proper equation of state. To describe the properties of crystalline silicon, the empirical Johnson-Cook (JC) material model was used. The detailed description of the model and the material parameters used in present simulations are presented in the Supplementary Information (SI).

In the simulation results presented below, the diameter of the laser irradiation spot on the sample surface was 25 µm ($1/e^2$), and the laser fluence was 1.7 J/cm$^2$ which closely corresponds to the experimental conditions. A possible ablation has been disregarded to evaluate only the mechanical response of the sample.

The simulation domain is presented in the bottom of figure 7. Its dimensions are 160×160×3 µm$^3$ along the ($x$, $y$, $z$) axes respectively. The central part of the domain of 20×20×3 µm$^3$ is zoomed to illustrate the model details. The discretization of the region is 1×1×0.075 µm$^3$ along the ($x$, $y$, $z$) axes respectively. The simulation region is discretized by 1030000 elements and 1063000 nodes. The sample was irradiated normally to the surface in the central part of the simulation domain by two successive pulses (their centers are marked by green dots; the pulse on the right comes first) with a time separation of 10 µs and a laser spot centers distance (LSCD) of 16 µm. To make the computation time reasonable, the time between the two pulses was shortened compared to that in our experiments (20 µs for figure 6). This is justified by the fact that, after 10 µs, the absorbed heat has substantially dissipated, and the laser-induced stress is finally imprinted to the sample. It must be noted that, when LSCD ≥ 25 µm, the developed stress and strain distributions are independent with no signs of interaction. The stress and strain amplifications are clearly achieved in simulations when LSCD is smaller than 25 µm, that is consistent with the experiments (see comments to figure 4). Thus, we analyse here the simulation results for the LSCD of 16 µm as a representative case.

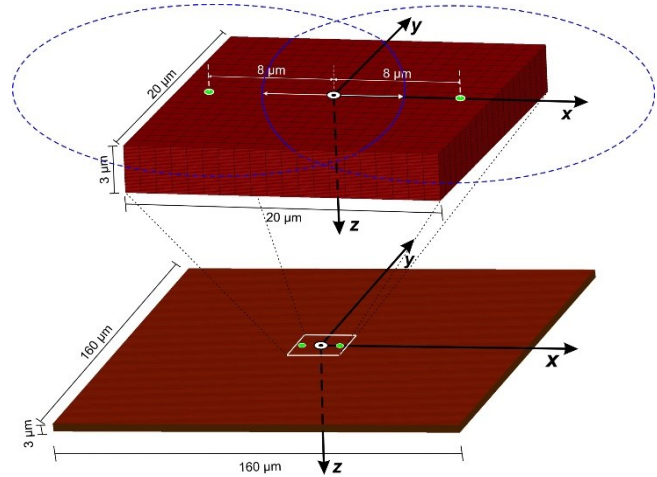

**Figure 7**. The scheme of the computational domain for modeling stress wave evolution upon irradiation by the two laser pulses in the middle of the sample, which are separated in time and space. The centers of the irradiation spots are shown by green dots. The sizes along the $x$ and $y$ directions are equal to 160 µm. The vertical dimension (along the $z$-axis perpendicular to the sample plane) is 3 µm. The zoomed part of the domain shown on the top demonstrates the laser irradiation spots (blue dash circles) on the sample surface (25 µm at $1/e^2$), LSCD of 16 µm, the orientation of the axes, and their origin. It illustrates the high resolution of the domain.

Figure 8 presents several characteristic snapshots of the effective stress (see details of the modeling in the SI) for the sample irradiation in the geometry shown in figure 7.

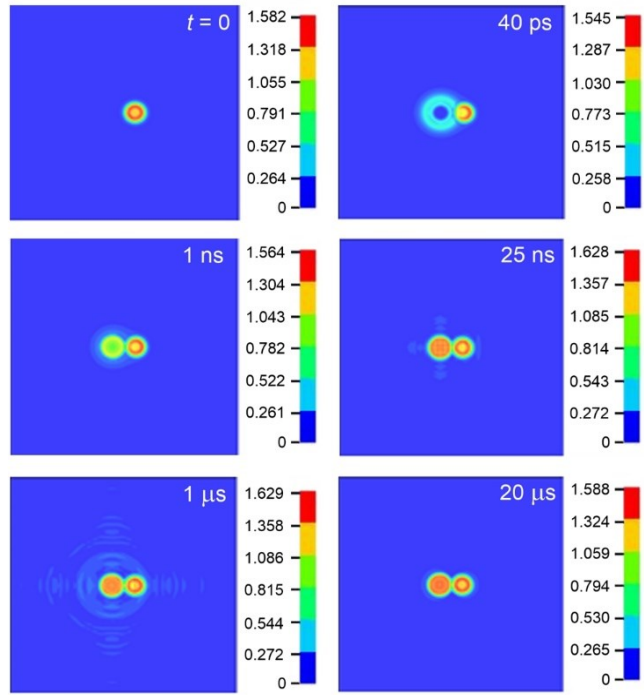

**Figure 8**. Snapshots of the effective stress (GPa) for the geometry shown in figure 7. The zero time corresponds to the moment when the second pulse couples the surface with the spot center of 16 μm to the left from the first spot center. The full computational frame of 160×160 μm$^2$ is shown. By $t$ = 20 μs, the residual stress distribution has been finally imprinted in the sample.

The residual stress from the first laser shot forms an annular structure with a maximum exceeding 1.5 GPa. The stress generated by the second pulse has first a ring shape of a sub-GPa level interacting with the structure generated by the first pulse. By a 1-ns time interval, this stressed region shrinks into a more uniform circle area, and the stress level reaches ~1 GPa. In the next 25 ns, the stress induced by the second pulse achieves its maximum value, which is preserved till about 1 μs. The emission of the surface acoustic waves is observed due to the stress concentration in this subdomain. It should be noted that the residual stress from the first laser pulse is not noticeably changing, although it affects the dynamics and the final stress level induced by the second laser pulse. Indeed, in the presence of the initial leading-pulse-generated stressed area, the stress from the second pulse is amplified by about 3% in the regime simulated. The snapshot at 20 μs corresponds to the stress finally imprinted to the sample. The stress appears in the form of a dimple (vertical displacement of the sample surface along $z$-axis) in the laser-affected area. The dynamics of the morphology of the stressed area is shown in figure 9 as the $z$-displacement distribution along the $x$-axis (see figure 7).

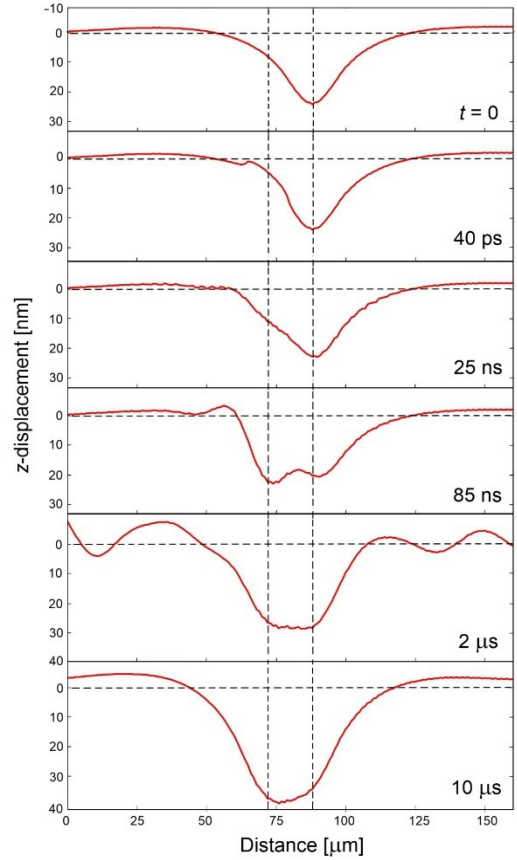

**Figure 9**. Profiles of the vertical displacement along the *x*-axis for the case shown in figure 8. The centers of the laser spots are marked by vertical dashed lines. The initial level of the surface is highlighted by the horizontal dashed lined.

The first laser pulse generates a dimple in the laser-affected zone with a depth of about 23 nm (figure 9, $t = 0$). The dimple is the result of a swift radial material expansion under the laser-induced stress. Similar dimples in semiconductor materials were observed under ion beam bombardment at low ion fluxes in the regimes below ablation/sputtering [55] (we note that the mechanisms of material modification by laser action and ion beam bombardment are similar [56]). According to the simulations, the laser-induced stress (figure 8) is much higher than the elastic limit of silicon (165-180 MPa [57]), thus leading to irreversible material deformation within the stressed zone. When the second pulse couples with the surface in the nearest proximity to the first-pulse-affected zone, even with a small overlap between the irradiation spots, its action first slightly diminishes the dimple depth while, at a nanosecond time scale, the *z*-displacement starts to grow with the signs of emitting acoustic waves (figures 8 and 9, 25 ns). The final dimple is merged from the action of the two laser pulses and reaches almost 40 nm. The corresponding maximum strain is evaluated as $\sim 5.4 \times 10^{-2}$. Moreover, the maximum computed temperatures during the two-shot irradiation are found to be ~2000 °C, which is much lower than the boiling point of Si (~3250 °C).

The considered simplified dynamics suggest that the residual stress induced on a sample surface by one laser pulse amplifies the stress and strain generated by the next laser pulse, even with a relatively low overlap between the irradiation spots. Considering the surface acoustic waves produced by the second (and following) laser pulse(s), the stress-induced changes of the band gap and the refractive index [58,59] should affect their propagation and plausibly contribute to the dynamics of the LIPSS formation. Thus, according to the Lorenz–Lorentz relation [60], the change of silicon density by 1% should result in the refractive index change of ~0.11, which is a considerable variation for the propagation of the electromagnetic waves (refractive index of silicon is ~3.5).

Although ablation and crater formation were not computationally included in this work, it can be anticipated that, additionally to the crater edge effect discussed in the previous sections, the laser-induced hydrodynamics with associated stress formation provide an additional factor for influencing surface nanostructuring that should be material-dependent. This influence is conditioned not only by hydrodynamic instabilities in the liquid phase [42-44] but also by the longer time effects due to imprinting strain-stress features after material resolidification that calls for further studies. Our future work will include the modelling of ablation and the simulation of crater formation, in relation to the hydrodynamic effects and the experimental material removal findings, presented in this work.

## 4. Conclusions

On the example of silicon, it has been demonstrated that, upon laser scanning at rather small overlapping between irradiation spots, regular LIPSS can be produced already starting from the second laser pulse, providing the polarization direction coincides with the scanning direction. If these directions are perpendicular to each other, LIPSS are not observed even after many pulses. This coherence effect, which was investigated using several irradiation patterns, is attributed to the scattering of surface electromagnetic waves from the edges of the crater created by a previous laser pulse. Thus, the overall mechanism can be seen as a three-wave interference: SEWs generated on the irradiated surface and SEWs scattered from the crater edge, which together interfere with the incoming laser pulse. The three-wave interference mechanism leads to the amplification of the periodic light absorption pattern that can be used for controlling the direct writing of the regular LIPSS.

Considering that the SEWs can also be scattered from the laser-modified regions, where the refractive index can be changed due to amorphization or residual stress formed by previous laser pulses, finite element simulations considering hydrodynamic effects have been performed to gain insight into possible consequences. Such kind of 3D modeling represents the first attempt to link the LIPSS formation with the modification fingerprints left by each successive laser pulse. The simulations have shown that both stress and $z$-displacement can be amplified upon laser scanning, thus plausibly enabling an additional degree of control over surface structuring that calls for further studies.

**Data availability statement**

All data that support the findings of this study are included within the article and in Supplementary Information.

**Acknowledgements**

I.M., Y.L., A.V.B. and N.M.B. are grateful for the support from Operational Programme Johannes Amos Comenius financed by European Structural and Investment Funds and the Czech Ministry of Education, Youth, and Sports (Project SENDISO No. CZ.02.01.01/00/22_008/0004596). V.D, H.P. and E.K. acknowledge the support with computational time granted by the Greek Research & Technology Network (GRNET) in the National HPC facility ARIS-under project ID pr016025-LaMPIOS III.

# Supplementary Information

**Mathematical model**

The one-temperature heat conduction equation, neglecting the convective and radiated energy transports, is written as follows

$$\rho(\mathbf{r},T)C_p(\mathbf{r},T)\frac{\partial T(\mathbf{r},T)}{\partial t} - \nabla[k(\mathbf{r},T)\nabla T(\mathbf{r},t)] = Q(\mathbf{r},t). \tag{S1}$$

Here $\rho$ is the mass density, $\mathbf{r}$ is the vector of location, $C_p$ is the specific heat at constant pressure, $T$ is the material temperature, $k$ is the thermal conductivity, $Q(\mathbf{r},t)$ is the absorbed laser energy per unit volume per unit time by the sample. The latent heat of melting or vaporization is also considered when temperature exceeds the melting or boiling point respectively.

Rapid local heating and associated thermal expansion generate a stress field, resulting in ultrasonic waves that propagate in the material. The equation of wave propagation is written in the form

$$\rho(\mathbf{r},T)\frac{\partial^2 U(\mathbf{r},T)}{\partial t^2} = \mu\nabla^2 U(\mathbf{r},T) + (\lambda + \mu)\nabla[\nabla U(\mathbf{r},t)] - \alpha(3\lambda + 2\mu)\nabla T(\mathbf{r},t) \tag{S2}$$

where $U$ is the displacement, $\lambda$ and $\mu$ are the material-dependent Lamé constants, and $\alpha$ is the thermoelastic expansion coefficient.

The mechanical behavior of the solid target is expressed by the equations for the stress-strain tensors as

$$\sigma_{ij} = 2\mu\varepsilon_{ij} + \lambda\varepsilon_{kk}\delta_{ij} - (3\lambda + 2\mu)\alpha(T-T_0)\delta_{ij} \tag{S3}$$

$$\varepsilon_{ij} = \frac{1}{2}\left(\frac{\partial U_i}{\partial x_j} + \frac{\partial U_j}{\partial x_i}\right) \tag{S4}$$

$T_0$ is the ambient temperature, $\sigma_{ij}$ and $\varepsilon_{ij}$ are the stress and strain tensors in the $i,j$-plane. The problem is solved in three dimensions.

The mathematical modeling of the hydrodynamic and the bulk behavior of the Si target due to the very high strain rates is performed by the Grüneisen Equation-of-State (EOS) for compressed materials:

$$p = \frac{\rho_0 C^2 \mu\left[1+\left(1-\frac{\gamma_0}{2}\right)\mu - \frac{\alpha_1}{2}\mu^2\right]}{[1-(s_1-1)\mu - s_2\frac{\mu^2}{\mu+1} - s_3\frac{\mu^3}{(\mu+1)^2}]^2} + (\gamma_0 + \alpha_1\mu)E \tag{S5}$$

For expanded materials results to:

$$p = \rho_0 C^2 \mu + (\gamma_0 + \alpha_1\mu)E \tag{S6}$$

Where, $E$ is the internal energy per initial volume, $C$ is the sound speed, $\mu$ is a volumetric parameter defined by the relation $\mu = (\rho/\rho_0) - 1$, and $\gamma_0$, $\alpha_1$, $s_1$, $s_2$, $s_3$ are unitless constants. $s_1$, $s_2$, and $s_3$ are the coefficients of the slope in the $u_p$–$u_s$ curve (the shock wave velocity $u_s$ varies linearly, with respect to the particle velocity $u_p$), $\gamma_0$ is the Grüneisen parameter and $\alpha_1$ represents the first order volume correction to $\gamma_0$ [1].

The laser pulse has the Gaussian temporal and spatial profiles, and the beam intensity of the beam propagating in the $y$ direction is expressed as

$$I(\mathbf{r},t) = I_0(t)e^{-4\ln 2\left(\frac{t-t_0}{t_0}\right)^2} e^{-\frac{x^2+y^2}{r_0^2}} \tag{S7}$$

with $r_0$ and $t_0$ to be the beam radius (at $1/e^2$ intensity level) and the FWHM laser pulse duration, respectively. $r$ is the distance from the center of the irradiation spot on the sample surface where the intensity is $I_0(t)$. Then, the *local absorbed energy* $Q(r,t)$ in the solid target is described as

$$Q(r,t) = I(r,t)(1-R)\alpha_b\, e^{-\alpha_b z} \tag{S8}$$

where $\alpha_b$ is the absorption coefficient of the material, and $R$ is the reflectivity of the surface.

**Laser beam parameters**

The beam diameter on the sample surface is assumed to be 25 µm (at $1/e^2$ intensity level). In all simulations, the FWHM pulse duration was 10 ps. The reflectivity of silicon at 1030 nm is taken for simplicity to be 0.32 at room temperature [2], and the absorption coefficient is temperature-dependent [3].

**Material properties**

To describe the properties of crystalline silicon, the empirical Johnson-Cook (JC) material model is used, which gives the flow stress $\sigma$ using the following equation

$$\sigma = (A + B\varepsilon^n)\left(1 + C \ln\frac{\dot{\varepsilon}}{\dot{\varepsilon}_0}\right)\left[1 - \left(\frac{T-T_0}{T_m - T_0}\right)^m\right] \tag{S9}$$

Here, $A$, $B$, $C$, $n$, and $m$ are experimental constants which depend on the material, $T_m$ is the melting point, $\dot{\varepsilon}$ and $\dot{\varepsilon}_0$ are the strain rate and the reference strain rate. In the case of high strain rates, the JC material model includes a fracture model that defines the equivalent plastic strain in case of damage. It is given by the following equation

$$\varepsilon_f = \left(D_1 + D_2 e^{D_3 \frac{p}{\sigma_{VM}}}\right)\left(1 + D_4 \ln\frac{\dot{\varepsilon}}{\dot{\varepsilon}_0}\right)\left(1 + D_5 \frac{T-T_r}{T_m - T_r}\right) \tag{S10}$$

where $p$ is the pressure, $D_1$, $D_2$, $D_3$, $D_4$, and $D_5$ are the failure parameters of the material, and $\sigma_{VM}$ is the Von Mises stress. The damage parameter D is defined as

$$D = \sum \frac{\Delta\varepsilon}{\varepsilon_f}. \tag{S11}$$

When the damage parameter $D$ becomes 1, the material fracture occurs. The properties used for crystalline silicon are given in tables S1-S5.

**Table S1.** Elastic properties of crystalline silicon [4].

| Density | 2330 kg/m³ |
|---|---|
| Young Modulus | 180 Gpa |
| Bulk Modulus | 135 Gpa |
| Shear Modulus | 60 Gpa |
| Poisson ratio | 0.28 |
| λ (1st Lame coefficient) | 89 Gpa |

| μ (2nd Lame coefficient) | 70 Gpa |

**Table S2.** JC parameters for crystalline silicon [4].

| A (initial yield of the material) | 0.896 GPa |
| B (strain hardening coefficient) | 0.530 GPa |
| n (strain hardening exponent) | 0.375 |
| C (strain rate sensitivity) | 0.424 |
| m (thermal softening exponent) | 1 |

**Table S3.** Failure model parameters for silicon [3].

| $D_1$ (fracture parameter) | 0.48 |
| $D_2$ (fracture parameter) | 0.4 |
| $D_3$ (fracture parameter) | -0.1 |
| $D_4$ (fracture parameter) | 0 |
| $D_5$ (fracture parameter) | 0 |

**Table S4.** Thermal properties of crystalline silicon [3].

| Specific heat | 710 J/(Kg °C) |
| Melting point | 1410 °C |
| Thermal conductivity (20 °C) | 1.56 W/(cm °C) |
| Solidus temperature | 940 °C |
| Liquidus temperature | 1410 °C |
| Latent heat | 1787000 J/kg |

**Table S5.** Temperature-dependent properties of crystalline silicon [4].

| Temperature (°C) | Thermal expansion coefficient (°C$^{-1}$) | Heat capacity (J/(Kg °C)) | Thermal conductivity (W/(cm °C)) |
| --- | --- | --- | --- |
| 0 | $1.5 \times 10^{-6}$ | 650 | 3.90 |
| 27 | $2.6 \times 10^{-6}$ | 710 | 1.56 |
| 227 | $3.8 \times 10^{-6}$ | 810 | 0.80 |
| 427 | $4.8 \times 10^{-6}$ | 870 | 0.50 |
| 627 | $5.3 \times 10^{-6}$ | 910 | 0.36 |
| 827 | $5.5 \times 10^{-6}$ | 935 | 0.30 |
| 1027 | $5.8 \times 10^{-6}$ | 950 | 0.25 |
| 1410 | $6.0 \times 10^{-6}$ | 980 | 0.22 |
| 2000 | $6.0 \times 10^{-6}$ | 950 | 0.20 |

**Dimensions and mesh discretization**

The dimensions of the simulation region are 160×160×3 µm$^3$ along the ($x$, $y$, $z$) axes, respectively. The discretization of the region is 1×1×0.075 µm$^3$ along the ($x$, $y$, $z$) axes, respectively. Thus, the simulation region consists of 1030000 elements and 1063000 nodes. The multiphysics simulations were performed by the LS-DYNA FEM MPP solvers [5].